\documentclass[journal]{IEEEtran}

\usepackage{amsmath,amssymb,amsfonts}
\usepackage{amsthm}
\usepackage{graphicx}
\usepackage{booktabs}
\usepackage{multirow}
\usepackage{algorithm}
\usepackage{algorithmic}
\usepackage{xcolor}
\usepackage{cite}
\usepackage{url}
\usepackage{xurl}
\usepackage{tikz}
\usetikzlibrary{positioning,fit,arrows.meta,calc}
\usepackage{hyperref}
\usepackage{braket}
\usepackage{placeins}
\usepackage{float}

\newtheorem{theorem}{Theorem}

\begin{document}

\title{Complementarity-Preserving Generative Theory for Multimodal ECG Synthesis: A Quantum-Inspired Approach}

\author{
    Timothy~Oladunni,~\IEEEmembership{}%
    \thanks{T. Oladunni, Farouk Ganiyu-Adewumi
, K. Maclin and C. Baidoo are with the Department of Computer Science, Morgan State University, Baltimore, MD, USA.
    (e-mails: timothy.oladunni@morgan.edu; fagan1@morgan.edu; kymac2@morgan.edu; clbai4@morgan.edu).}
    Farouk Ganiyu-Adewumi,
    Clyde Baidoo,
    and Kyndal Maclin
}

\maketitle

\begin{abstract}
Multimodal deep learning has substantially improved electrocardiogram (ECG) classification by jointly leveraging time, frequency, and time–frequency representations. However, existing generative models typically synthesize these modalities independently, resulting in synthetic ECG data that are visually plausible yet physiologically inconsistent across domains. This work establishes a Complementarity-Preserving Generative Theory (CPGT), which posits that physiologically valid multimodal signal generation requires explicit preservation of cross-domain complementarity rather than loosely coupled modality synthesis.
We instantiate CPGT through Q–CFD–GAN, a quantum-inspired generative framework that models multimodal ECG structure within a complex-valued latent space and enforces complementarity-aware constraints regulating mutual information, redundancy, and morphological coherence. Experimental evaluation demonstrates that Q–CFD–GAN reduces latent embedding variance by 82\%, decreases classifier-based plausibility error by 26.6\%, and restores tri-domain complementarity from 0.56 to 0.91, while achieving the lowest observed morphology deviation (3.8\%).
These findings show that preserving multimodal information geometry, rather than optimizing modality-specific fidelity alone, is essential for generating synthetic ECG signals that remain physiologically meaningful and suitable for downstream clinical machine-learning applications.
\end{abstract}


\IEEEpeerreviewmaketitle

\section{Introduction}

Electrocardiography (ECG) is a widely used diagnostic tool in cardiovascular
medicine due to its low cost, broad accessibility, and ability to capture subtle
electrophysiological abnormalities. Despite substantial progress in deep
learning–based ECG analysis, model performance remains constrained by the
limited availability and class imbalance of clinically meaningful data. Many
high-risk conditions, including acute ischemia, conduction abnormalities, and
malignant arrhythmias, occur infrequently in real-world datasets, limiting the
generalizability of supervised models, particularly in low-resource and
underrepresented clinical settings.

Generative adversarial networks (GANs) have been explored as a strategy for
mitigating data scarcity by synthesizing ECG signals for data augmentation,
simulation, and privacy-preserving data sharing. However, most existing ECG
generative models synthesize individual representations: time-domain waveforms,
frequency-domain spectra, or time-frequency decompositions, largely
independently. This design choice neglects the coordinated relationships among
modalities imposed by cardiac electrophysiology. As a result, synthetic ECG
signals may appear plausible in one domain while exhibiting inconsistent
spectral structure or degenerate time--frequency patterns in others. Such
cross-domain incoherence substantially limits the usefulness of synthetic data
for modern multimodal classifiers that rely on consistent evidence across
representations.

While multimodality is often defined in terms of heterogeneous data sources or multiple sensing modalities, this work adopts a signal-domain interpretation of multimodality for ECG analysis. An ECG signal can be represented in multiple domains, including the time, frequency, and time–frequency domains, each capturing distinct physiological characteristics of cardiac activity. The time domain preserves waveform morphology such as the P wave, QRS complex, and T wave, the frequency domain highlights spectral properties associated with rhythm and periodicity, and the time–frequency domain captures transient behaviors that evolve over time. Because these representations encode complementary information about the same physiological process, this work treats each signal domain as a distinct modality within a unified multimodal learning framework.

\textbf{The overarching goal of this study is to resolve the physiological inconsistencies inherent in current  generative models for electrocardiogram (ECG) synthesis by shifting the modeling paradigm from independent modality optimization to the preservation of cross-domain information geometry.} To achieve this, we formulate the \textit{Complementarity-Preserving Generative Theory} (CPGT), which establishes the mathematical requirements for maintaining coupled joint structures across time, frequency, and time--frequency representations. By instantiating this theory through a quantum-inspired generative framework (\textit{Q--CFD--GAN}), we aim to demonstrate that explicitly enforcing complementarity-aware constraints within a complex-valued latent space significantly reduces morphology deviation and latent manifold drift. Ultimately, \textit{this work seeks to provide a theoretically grounded methodology, backed by empirical evidence, for generating high-fidelity synthetic ECG signals that remain physiologically meaningful and suitable for robust deployment in downstream clinical machine-learning applications.}

To clarify the objectives of the empirical evaluation, Table~\ref{tab:rq_mapping} summarizes the key research questions addressed by the quantitative experiments and maps them to the corresponding tables and figures in this study. The evaluation is designed to examine three fundamental aspects of multimodal ECG generation: statistical fidelity to real signals, preservation of cross-domain complementarity, and physiological plausibility of the synthesized signals. Table~II evaluates statistical and complementarity-related metrics, while Figures~3--6 analyze morphological, spectral, and energy-related characteristics of the generated ECG signals. Together, these experiments provide a comprehensive assessment of whether Q--CFD--GAN preserves both the statistical structure and physiologically meaningful relationships present in real multimodal ECG data.

\begin{table*}[t]
\centering
\caption{Research Questions Addressed by Quantitative Evaluation}
\label{tab:rq_mapping}
\renewcommand{\arraystretch}{1.2}
\begin{tabular}{p{1.8cm} p{3.6cm} p{2.2cm} p{5.2cm}}
\toprule
\textbf{RQ} & \textbf{Objective} & \textbf{Evidence} & \textbf{Key Finding} \\
\midrule

RQ1 &
Does Q--CFD--GAN preserve the statistical structure of real ECG signals better than baseline generative models? &
Table II ($\sigma^2$, $\Delta$, $\rho_\Delta$) &
Q--CFD--GAN achieves the lowest variance deviation ($\sigma^2=0.033$) and smallest structural error ($\Delta=0.157$), indicating improved statistical fidelity. \\

RQ2 &
Does Q--CFD--GAN better preserve multimodal complementarity across ECG domains? &
Table II ($\widehat{C}_{TFS}/C_{TFS}$) &
Q--CFD--GAN restores 91\% of real multimodal complementarity, outperforming MI--GAN (0.56) and Q--Base (0.68). \\

RQ3 &
Does interference-aware latent modeling improve multimodal synthesis quality? &
Table II ($\gamma$, $\kappa$) &
Q--CFD--GAN achieves the highest constructive interference score ($\gamma=1.625$) and lowest redundancy coefficient ($\kappa=0.339$). \\

RQ4 &
Do generated ECG signals preserve physiologically meaningful amplitude characteristics? &
Fig.~3 &
The peak-to-peak amplitude distribution of CFD-Syn closely matches real ECG signals, while NonCFD-Syn collapses to unrealistic ranges. \\

RQ5 &
Do generated signals maintain realistic spectral complexity? &
Fig.~5 &
CFD-Syn preserves spectral entropy characteristics of real ECG signals, while NonCFD-Syn shows entropy collapse. \\

RQ6 &
Do generated signals maintain realistic signal energy characteristics? &
Fig.~4 &
CFD-Syn follows the RMS energy distribution of real ECG signals, indicating physiologically plausible signal energy levels. \\

RQ7 &
Does the model preserve physiologically consistent relationships between signal amplitude and energy? &
Fig.~6 &
Q--CFD--GAN preserves the joint RMS--P2P manifold observed in real ECG signals, while NonCFD-Syn collapses into a degenerate region. \\

\bottomrule
\end{tabular}
\end{table*}

\subsection{Contributions}
The primary contribution of this work is the shift from modality-independent generative modeling to a theory-grounded, complementarity-preserving framework for multimodal signal synthesis. The specific contributions of this article are fourfold:

\begin{itemize}
    \item \textbf{Establishment of Complementarity-Preserving Generative Theory (CPGT):} We propose a novel theoretical framework grounded in statistical mechanics and information theory. CPGT posits that the synthesis of multimodal signals, such as ECG, requires the explicit preservation of the coupled joint structure (information geometry) between representations rather than the independent optimization of unimodal fidelity.
    
    \item \textbf{Introduction of the Q--CFD--GAN Framework:} We instantiate CPGT through a quantum-inspired generative adversarial network ($Q$--$CFD$--$GAN$). This architecture utilizes a complex-valued latent space to model cross-domain interference and enforces a set of ``complementarity-aware'' constraints that regulate mutual information, redundancy, and morphological stability across time, frequency, and time--frequency domains.
    
    \item \textbf{Rigorous Empirical Validation of Physiological Fidelity:} Through extensive evaluation against state-of-the-art baselines (MI-GAN and Q-Base), we demonstrate that $Q$--$CFD$--$GAN$ restores tri-domain complementarity from $0.56$ to $0.91$. Furthermore, the framework achieves a $26.6\%$ reduction in classifier-based plausibility error and a $3.8\%$ morphology deviation, significantly outperforming traditional generative approaches.
    
    \item \textbf{Demonstration of Clinical and Translational Utility:} We provide evidence that preserving multimodal information geometry is essential for generating synthetic data suitable for downstream clinical machine-learning applications. We demonstrate that our approach mitigates manifold drift (reducing latent variance by $82\%$), ensuring synthetic samples remain within the ``admissible manifold'' of real-world physiological signals.
\end{itemize}
\noindent 

The remainder of the paper is organized as follows.
Section~II reviews prior work on ECG generation, multimodal ECG analysis, and
related representational approaches. Section~III introduces the theoretical
framework underlying Q--CFD--GAN, including CFD-based regularization, the
interference-aware latent representation, and the cross-domain coupling
mechanism. Section~IV details preprocessing and architectural design.
Section~V outlines the experimental setup and baselines. Section~VI presents
empirical results and ablation studies, followed by discussion and limitations
in Section~VII. Section~VIII concludes the paper.

\section{Related Work}

\subsection{ECG Generative Models}
Most existing ECG generative models operate exclusively in the time domain,
synthesizing waveform segments that approximate P--QRS--T morphology.
WaveGAN-style architectures~\cite{donahue2019adversarial} and temporal models
such as TimeGAN~\cite{yoon2019time} demonstrate that adversarial learning can
capture coarse temporal dynamics for applications including arrhythmia data
augmentation. Simulator-assisted approaches, including SimGANs~\cite{golany2020simgans},
introduce physiology-informed priors to improve waveform realism. More recently,
diffusion-based methods such as DiffECG~\cite{neifar2023diffecg} and structured
conditional diffusion models~\cite{alcaraz2023sssd} have improved sampling
stability relative to GAN-based approaches.

Despite these advances, most ECG generative frameworks synthesize a \emph{single}
representation, typically the raw waveform, and do not enforce consistency across
time, frequency, and time-frequency domains. Even when auxiliary
representations are derived post hoc, no principled mechanism ensures alignment
between spectral structure, temporal morphology, and joint time--frequency
patterns. As a result, synthetic ECG data may exhibit plausible morphology in
one domain while producing inconsistent or degenerate representations in
others. Such cross-domain incoherence limits the usefulness of synthetic data
for modern multimodal classifiers that rely on coordinated evidence across
representations.

\subsection{Multimodal ECG Analysis and Complementarity}
A growing body of work has demonstrated that heterogeneous ECG representations
improve diagnostic robustness and interpretability. Studies based on large-scale
datasets such as PTB-XL~\cite{strodthoff2021deep} show that spectral and
time--frequency representations capture diagnostic cues that are not fully
expressed in raw waveforms. Hybrid architectures combining convolutional,
recurrent, and attention-based models further highlight the benefits of fusing
multiple representations and auxiliary clinical variables for arrhythmia
detection and risk prediction~\cite{tawfeek_2025_cardiovascular,yildirim2018ecg}.

Complementary Fusion Domain (CFD) theory posits that multimodal ECG performance is
governed by the \emph{complementarity} of feature domains rather than by the
number of domains fused. This theory established that time (T), frequency (F),
and time--frequency (S) representations encode distinct, non-redundant
diagnostic information whose relationships critically influence learning
outcomes \cite{oladunni2025cfd},\cite{oladunni2025yale}. In subsequent work, this principle was formalized through quantitative
measures of synergy, redundancy, and robustness to adversarial attacks, providing an
information-theoretic framework for analyzing multimodal ECG representations \cite{11239065}.
However, existing generative models do not attempt to preserve this
complementarity during synthesis, instead treating multimodal outputs as
loosely coupled signals rather than a coordinated information-theoretic object.

\subsection{Complex-Valued and Quantum-Inspired Representation Learning}
Complex-valued and quantum-inspired learning frameworks provide mathematical
tools for modeling amplitude--phase interactions, superposition, and
interference~\cite{lai2023quantum}. Hilbert-space embeddings and
complex-valued representations have been explored for supervised and
unsupervised learning~\cite{wiebe2014quantum}, natural language modeling~\cite{zhao2022quantumnlp},
and general signal analysis~\cite{schuld2015introduction}. These approaches are
particularly well suited to oscillatory signals, where phase relationships and
interference patterns play a central role.

In the context of ECG, depolarization and repolarization processes generate
oscillatory waveforms whose morphology depends on both amplitude and phase
relationships across time and frequency. While quantum-inspired representations
offer a natural mechanism for modeling such interactions, prior work has focused
primarily on representational capacity rather than physiological validity. No
existing framework jointly (i) models phase-dependent interactions, (ii)
preserves multimodal ECG complementarity, and (iii) enforces morphology-aware
constraints during signal generation.

Table~\ref{tab:comparison} summarizes the limitations of prior approaches and
highlights how Q--CFD--GAN addresses this gap by integrating complex-valued,
interference-aware latent modeling with complementarity-preserving
regularization and physiology-aware loss functions.

\vspace{0.25em}
\vspace{0.25em}
\begin{table}[t]
\centering
\caption{Summary Comparison of Q--CFD--GAN with Prior Frameworks}
\label{tab:comparison}
\begin{tabular}{p{2.3cm} p{5.7cm}}
\toprule
\textbf{Framework} &
\textbf{Key Limitation for Multimodal ECG Generation} \\
\midrule

Classical GANs~\cite{donahue2019adversarial,yoon2019time,golany2020simgans} &
Single-domain generation with no mechanism for synchronized tri-domain
(T/F/S) coherence. \\[0.5em]

InfoGAN and MI-based GANs~\cite{chen2016infogan} &
Encourage disentanglement but do not regulate redundancy, orthogonality, or
ECG-specific complementarity. \\[0.8em]

Multimodal ECG classifiers~\cite{strodthoff2021deep,tawfeek_2025_cardiovascular,oladunni2026modelchangemindenergybased,yildirim2018ecg} &
Discriminative models only; incapable of synthesizing coherent multimodal ECG
signals. \\[0.8em]

Complementarity-based analysis frameworks &
Quantify multimodal relationships but do not enforce complementarity during
data generation. \\[0.8em]

Quantum-inspired models~\cite{schuld2015introduction,wiebe2014quantum,zhao2022quantumnlp} &
Capture phase and interference effects but lack physiology-aware constraints and
ECG-specific multimodal structure. \\[0.8em]

Cardiac simulation models~\cite{franzone2014mathematical,niederer2019computational} &
Biophysically detailed but rigid; not aligned with learned information geometry
or data-driven multimodal coupling. \\[0.8em]

\textbf{Q--CFD--GAN (This work)} &
First framework to integrate complex-valued latent interference,
complementarity-preserving regularization, and morphology-aware constraints to
generate physiologically coherent, synchronized tri-domain ECG signals. \\

\bottomrule
\end{tabular}
\end{table}

\section{Q--CFD--GAN Theory}

This section presents the theoretical foundations of Q--CFD--GAN. Our analysis focuses on representational structure and coupling
constraints enforced by the proposed formulation, independent of the specific
optimization dynamics used to train the model.

\subsection{Complementary Feature Domain Structure}

Let $T$, $F$, and $S$ denote the time-domain waveform, frequency-domain magnitude
spectrum, and time-frequency scalogram of an ECG segment, respectively. Per CFD principle, these representations encode partially overlapping yet
non-redundant diagnostic information, and their joint complementarity
governs the effectiveness of multimodal ECG analysis \cite{11239065}.

\textbf{Modality informativeness.}
For each modality $X \in \{T, F, S\}$, informativeness with respect to the class
label $Y$ is quantified using mutual information:
\begin{equation}
I_X = I(X; Y).
\end{equation}

\textbf{Pairwise redundancy.}
The redundancy between two modalities $X$ and $Z$ is defined as
\begin{equation}
R_{XZ} = I(X; Y) + I(Z; Y) - I(X, Z; Y),
\end{equation}
which captures the extent to which the modalities provide overlapping diagnostic
evidence.

\textbf{Tri-domain complementarity.}
Overall multimodal complementarity is expressed as
\begin{equation}
C_{TFS} = I_T + I_F + I_S
- R_{TF} - R_{FS} - R_{ST}.
\end{equation}

A generative model is said to be \emph{CFD-preserving} if the complementarity and
redundancy statistics of the generated data approximate those observed in real
ECG recordings:
\begin{equation}
\widehat{C}_{TFS} \approx C_{TFS}, \qquad
\widehat{R}_{ij} \approx R_{ij}.
\end{equation}

These conditions formalize the requirement that synthetic multimodal ECG signals
retain the information-theoretic organization inherent to real physiological
data, rather than reproducing plausible but independently synthesized
modalities.

\subsection{Structured Latent Representation for Multimodal Coupling}

Preserving CFD structure during generation requires a latent representation that
can model phase-dependent interactions and regulated coupling across modalities.
While real-valued latent spaces can represent multimodal features, they do not
naturally encode relative phase relationships or interference effects without
additional constraints, which may lead to redundancy or modal dominance.

To address this, Q--CFD--GAN employs a structured complex-valued latent
representation defined in a Hilbert space:
\begin{equation}
|\Psi\rangle =
\alpha_T |T\rangle + \alpha_F |F\rangle + \alpha_S |S\rangle,
\end{equation}
subject to the normalization constraint
\begin{equation}
|\alpha_T|^2 + |\alpha_F|^2 + |\alpha_S|^2 = 1,
\qquad \alpha_i \in \mathbb{C}.
\end{equation}

Here, $\{|T\rangle, |F\rangle, |S\rangle\}$ form an orthonormal basis associated
with the time, frequency, and time--frequency ECG domains. This formulation
introduces a structured inductive bias that supports phase-sensitive
cross-domain interactions and controlled interference among modalities.

Although an equivalent real-valued embedding could, in principle, encode the
same degrees of freedom, complex-valued representations naturally impose
phase-coherent coupling constraints that would otherwise require explicit and
potentially unstable regularization. In this sense, the complex formulation
provides a minimal and well-conditioned mechanism for enforcing multimodal
complementarity during synthesis.

\subsection{Interference Operator and Cross-Domain Regulation}

Cross-domain interactions within the latent space are regulated by a Hermitian
interference operator
\begin{equation}
\hat{H}_{\mathrm{int}} =
\begin{bmatrix}
0 & w_{TF} & w_{TS} \\
w_{FT} & 0 & w_{FS} \\
w_{ST} & w_{SF} & 0
\end{bmatrix},
\end{equation}
where $w_{ij}$ controls the coupling strength between modalities $i$ and $j$.
Hermiticity ensures symmetric interactions and guarantees that the resulting
interference energies are real-valued.

The interference energy associated with a latent ECG state $|\Psi\rangle$ is
defined as
\begin{equation}
\mathcal{I}(\Psi) = \langle \Psi | \hat{H}_{\mathrm{int}} | \Psi \rangle \in
\mathbb{R}.
\end{equation}

This quantity regulates constructive and destructive interactions across
modalities, preventing uncontrolled dominance or collapse of any single
representation.

\paragraph{Interference collapse.}
In the absence of regulated coupling, multimodal generators tend to favor the
most easily synthesized modality, causing other representations to become weakly
informative or redundant. In ECG synthesis, this failure mode manifests as signals
that appear morphologically plausible in the time domain while exhibiting
inconsistent spectral structure or degenerate time--frequency patterns. The
interference operator mitigates this behavior by explicitly coupling modalities
in the latent space, enforcing coordinated multimodal structure during
generation.

\subsection{Complementarity and Morphology Constraints}

\textbf{CFD-preserving constraint.}
To enforce preservation of multimodal complementarity during synthesis, we
define the CFD loss as
\begin{equation}
\mathcal{L}_{\mathrm{CFD}} =
\left\| C_{TFS} - \widehat{C}_{TFS} \right\|
+ \lambda_{\mathrm{orth}} \sum_{i \neq j}
\left\| \langle X_i, X_j \rangle \right\|^2,
\end{equation}
where the first term aligns real and synthetic tri-domain complementarity, and
the second penalizes excessive cross-domain correlation, promoting orthogonality
consistent with CFD theory.

\textbf{Morphological constraint.}
Physiological structure is enforced by constraining clinically salient waveform
components:
\begin{equation}
\mathcal{L}_{\mathrm{phys}} =
\left\| \mathrm{QRS}_{\mathrm{real}} - \widehat{\mathrm{QRS}} \right\|_2^2
+
\left\| \mathrm{ST}_{\mathrm{real}} - \widehat{\mathrm{ST}} \right\|_2^2.
\end{equation}

\textbf{Unified generator objective.}
The generator is optimized using the composite objective
\begin{equation}
\mathcal{L}_{\mathrm{QCFD}} =
\mathcal{L}_{\mathrm{GAN}}
+ \lambda_1 \mathcal{L}_{\mathrm{CFD}}
+ \lambda_2 \mathcal{L}_{\mathrm{interf}}
+ \lambda_3 \mathcal{L}_{\mathrm{phys}}.
\end{equation}

\begin{algorithm}[!t]
\caption{Q--CFD--GAN Training Procedure}
\label{alg:q_cfd_gan}
\begin{algorithmic}[1]

\REQUIRE Real multimodal ECG dataset $\mathcal{D}$, complex latent prior $p(\Psi)$,
Hermitian interference operator $\hat{H}_{\mathrm{int}}$, CFD statistics
$(C_{TFS}, R_{ij})$, morphology priors, weighting coefficients $\lambda$.

\STATE Initialize generator $G$ and discriminator $D$.

\WHILE{not converged}
    \STATE Sample real batch $(T,F,S) \sim \mathcal{D}$.
    \STATE Sample complex latent states $|\Psi\rangle \sim p(\Psi)$ and normalize.
    \STATE Generate synthetic modalities
           $(\widehat{T}, \widehat{F}, \widehat{S}) = G(|\Psi\rangle)$.
    \STATE Apply interference-aware coupling:
           $\tilde{h} = h + \hat{H}_{\mathrm{int}} h$.
    \STATE Compute generator loss
    \[
       \mathcal{L}_G =
       \mathcal{L}_{\mathrm{GAN}}
       + \lambda_1 \mathcal{L}_{\mathrm{CFD}}
       + \lambda_2 \mathcal{L}_{\mathrm{interf}}
       + \lambda_3 \mathcal{L}_{\mathrm{phys}} .
    \]
    \STATE Update discriminator $D$ using WGAN-GP.
    \STATE Update generator $G$ using $\nabla_{\theta_G} \mathcal{L}_G$.
\ENDWHILE

\RETURN Trained generator $G$ (Q--CFD--GAN)

\end{algorithmic}
\end{algorithm}

\section{Proposed Q--CFD--GAN Framework}

The Q--CFD--GAN framework integrates three core components: a complex-valued
latent representation, an interference-aware coupling operator, and a
tri-domain decoder. Together, these components enable physiologically consistent
multimodal ECG generations by enforcing cross-domain complementarity,
interference stability, and morphology preservation.

\subsection*{Complex-Valued Latent Representation}

Each ECG sample is mapped to a complex-valued latent state
\[
\ket{\Psi} =
\alpha_T \ket{T} + \alpha_F \ket{F} + \alpha_S \ket{S},
\]
where $\{\ket{T}, \ket{F}, \ket{S}\}$ denote orthonormal basis states associated
with the time, frequency, and time--frequency domains. The complex amplitudes
$\alpha_i$ encode relative domain contributions and phase relationships,
providing a representational mechanism for capturing multimodal complementarity.

\subsection*{Interference-Aware Coupling Operator}

Cross-domain interactions are regulated by a Hermitian interference operator
$\hat{H}_{\mathrm{int}}$, which assigns each latent state an interference energy
\begin{equation}
\mathcal{I}(\Psi) = \bra{\Psi} \hat{H}_{\mathrm{int}} \ket{\Psi}.
\label{eq:interference_energy}
\end{equation}

Real ECG signals occupy stable interference-energy regimes reflecting
physiological coupling among domains. To enforce consistency, we define the
interference-preserving loss
\begin{equation}
\mathcal{L}_{\mathrm{interf}}
=
\left| \mathcal{I}(\Psi_{\mathrm{real}}) -
       \mathcal{I}(\Psi_{\mathrm{gen}}) \right|.
\label{eq:interference_loss}
\end{equation}

This term prevents uncontrolled dominance, collapse, or redundancy in the
latent superposition.

\subsection*{Tri-Domain Decoder}

The tri-domain decoder $(G_T, G_F, G_S)$ maps the latent state $\ket{\Psi}$ to
synthetic time-, frequency-, and time--frequency-domain representations. Each
branch specializes in reconstructing its corresponding modality while
collectively preserving the cross-domain structure encoded in the latent state.

\subsection*{Unified Q--CFD Objective}

The complete training objective is given by
\begin{equation}
\mathcal{L}_{\mathrm{QCFD}} =
\mathcal{L}_{\mathrm{CFD}} +
\mathcal{L}_{\mathrm{interf}} +
\mathcal{L}_{\mathrm{phys}} .
\label{eq:qcfdtotalloss}
\end{equation}

\noindent
\textit{Complementarity preservation.}
$\mathcal{L}_{\mathrm{CFD}}$ enforces retention of multimodal complementarity in
accordance with CFD theory.

\noindent
\textit{Physiological consistency.}
$\mathcal{L}_{\mathrm{phys}}$ penalizes deviations from clinically meaningful
ECG morphology.

\subsection*{Plausibility Gap Metric}

Clinical plausibility is quantified using the plausibility gap
\begin{equation}
\Delta_{\mathrm{model}} =
D_{\mathrm{KL}}
\!\left(
C(x_{\mathrm{real}})
\;\Vert\;
C(x_{\mathrm{gen}})
\right),
\label{eq:plausibility_gap}
\end{equation}
where $C(\cdot)$ denotes a pretrained ECG classifier. A small gap indicates that
synthetic ECG signals are indistinguishable from real signals in classifier
space.

\subsection*{Theoretical Guarantees}

The following results formalize conditions under which Q--CFD--GAN preserves
multimodal complementarity, interference stability, and tri-domain coherence.

\subsubsection*{Complementarity Preservation}

\begin{theorem}[CFD Preservation]
Let $\mathcal{L}_{\mathrm{CFD}} = 0$ iff
$(\widehat{C}_{TFS} = C_{TFS},\ \widehat{R}_{ij} = R_{ij})$.
If $G^\star = \arg\min_G \mathcal{L}_{\mathrm{CFD}}$, then
\[
\nabla_G \mathcal{L}_{\mathrm{CFD}}(G^\star) = 0,
\qquad
\widehat{C}_{TFS}(G^\star) = C_{TFS}.
\]
\end{theorem}

\noindent\textit{Interpretation.}
At a global minimizer of the CFD loss, synthetic data match the true multimodal
complementarity structure.

\subsubsection*{Minimal Complex Representation}

\begin{theorem}[Minimal Complex Latent Structure]
Preserving phase-dependent cross-domain interactions requires a latent
representation in $\mathbb{C}^3$; equivalently,
\[
\mathbb{C}^3 \simeq \mathbb{R}^6
\]
is the minimal real embedding capable of expressing multimodal interference.
\end{theorem}

\noindent\textit{Interpretation.}
Complex-valued representations provide the minimal structure necessary for
phase-coherent multimodal coupling.

\subsubsection*{Interference Stability}

\begin{theorem}[Interference Stability]
Let $\mathcal{I}(\Psi) = \langle \Psi | \hat{H}_{\mathrm{int}} | \Psi \rangle$.
If $\nabla_\Psi \mathcal{I}(\Psi^\star) = 0$ and $\widehat{R}_{ij}$ depend
smoothly on the coupling parameters, then
\[
|\widehat{R}_{ij}(\Psi^\star) - R_{ij}| \le \epsilon
\quad \Rightarrow \quad
\widehat{C}_{TFS}(\Psi^\star) \in \mathcal{N}(C_{TFS}).
\]
\end{theorem}

\noindent\textit{Interpretation.}
Stable interference equilibria prevent amplification of cross-domain redundancy.

\subsubsection*{Tri-Domain Coherence}

\begin{theorem}[Sufficient Conditions for Coherent Generation]
If
\[
\mathcal{L}_{\mathrm{CFD}}(G) = 0, \qquad
\nabla_\Psi \mathcal{I}(\Psi) = 0, \qquad
\nabla_G \mathcal{L}_{\mathrm{phys}}(G) = 0,
\]
then the generated modalities satisfy
\[
(\widehat{T}, \widehat{F}, \widehat{S})
\longrightarrow \mathcal{M}_{\mathrm{ECG}},
\]
the manifold of physiologically valid ECG signals.
\end{theorem}

\noindent\textit{Interpretation.}
Complementarity preservation, interference stability, and morphology
consistency jointly ensure physiologically coherent tri-domain generation.

\subsection*{Scientific Rationale}

Real ECG signals exhibit multimodal consistency, positive complementarity, and
phase-sensitive morphology. These properties imply that a generative model must:
\begin{itemize}
    \item employ a complex-valued latent representation to encode phase structure;
    \item regulate cross-domain interactions via a symmetric (Hermitian) operator;
    \item preserve multimodal complementarity through $\mathcal{L}_{\mathrm{CFD}}$;
    \item enforce physiological morphology via $\mathcal{L}_{\mathrm{phys}}$.
\end{itemize}

These requirements directly motivate the design of the Q--CFD--GAN framework.

\subsection{Multimodal Generator Architecture}

The generator $G$ maps a shared latent representation to three synchronized ECG
modalities corresponding to the time, frequency, and time--frequency domains:
\[
(\widehat{T}, \widehat{F}, \widehat{S}) = G(|\Psi\rangle).
\]
Unlike conventional multimodal GANs that generate each modality independently,
Q--CFD--GAN enforces shared structure prior to decoding, ensuring that all outputs
remain mutually consistent.

\begin{figure}[t]
\centering
\resizebox{\columnwidth}{!}{%
\begin{tikzpicture}[
    box/.style={
        rectangle, draw, rounded corners,
        minimum width=42mm,
        minimum height=11mm,
        align=center,
        font=\small
    },
    arrow/.style={-Latex, semithick},
    node distance=14mm and 20mm,
    >=latex
]

\node[box, fill=gray!10] (latent)
    {Complex-valued latent state $|\Psi\rangle$};

\node[box, below=16mm of latent] (core)
    {Shared latent core};

\node[coordinate, below=12mm of core] (split) {};

\draw[arrow] (latent.south) -- (core.north);
\draw[arrow] (core.south)  -- (split);

\node[box, below=18mm of split] (decF)
    {Frequency-domain decoder\\(Transformer)};
\node[box, left=22mm of decF] (decT)
    {Time-domain decoder\\(1D transposed CNN)};
\node[box, right=22mm of decF] (decS)
    {Time--frequency decoder\\(2D deconv CNN)};

\draw[arrow] (split) -- (decT.north);
\draw[arrow] (split) -- (decF.north);
\draw[arrow] (split) -- (decS.north);

\node[box, below=16mm of decT] (outT)
    {$\widehat{T}$\\Time-domain ECG};
\node[box, below=16mm of decF] (outF)
    {$\widehat{F}$\\Frequency representation};
\node[box, below=16mm of decS] (outS)
    {$\widehat{S}$\\Time--frequency scalogram};

\draw[arrow] (decT.south) -- (outT.north);
\draw[arrow] (decF.south) -- (outF.north);
\draw[arrow] (decS.south) -- (outS.north);

\end{tikzpicture}
}
\caption{Q--CFD--GAN generator architecture. A shared complex-valued latent state
is mapped to a common latent core and decoded in parallel to generate
synchronized time-, frequency-, and time--frequency ECG representations.}
\label{fig:generator}
\end{figure}
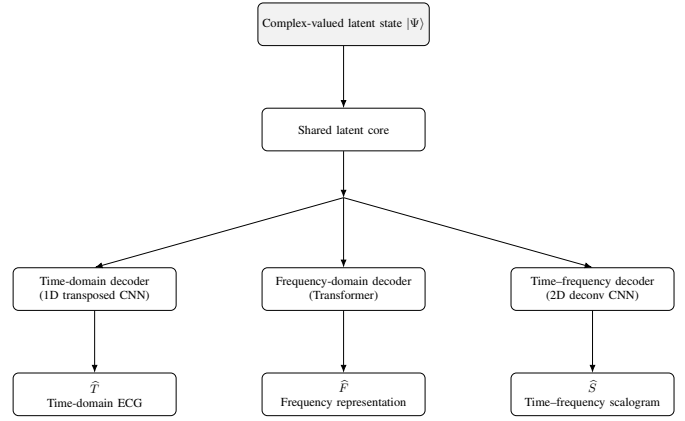

Figure~\ref{fig:generator} illustrates the generator architecture. A
complex-valued latent state is first mapped to a shared latent core that encodes
cross-domain dependencies. This shared representation is then decoded in
parallel by modality-specific generators. The design explicitly prevents
modality collapse by coupling the decoding process through a common latent
source.

\subsection{Multimodal ECG Preprocessing}

Each ECG beat is transformed into three complementary representations derived
from the same underlying signal, ensuring strict alignment across modalities.

\subsubsection{Time-domain signal ($T$)}
Raw ECG waveforms are bandpass-filtered (0.5--40\,Hz), baseline-corrected, and
amplitude-normalized. Each segment is padded or truncated to a fixed length of
$N=1000$ samples.

\subsubsection{Frequency-domain spectrum ($F$)}
The short-time Fourier transform is computed using a 256-point Hann window with
50\% overlap. The log-scaled magnitude spectrum is truncated to the first $K$
frequency bins.

\subsubsection{Time--frequency scalogram ($S$)}
A continuous wavelet transform with a Morlet basis produces a scalogram of size
$H \times W$, which is min--max normalized.

\subsection{Latent Sampling and Modality-Specific Decoding}

The generator samples a complex latent vector
\[
|\Psi\rangle = z_r + i z_i \in \mathbb{C}^d,
\]
where $z_r, z_i \sim \mathcal{N}(0,I)$. An $\ell_2$ normalization enforces
$\sum_k |\Psi_k|^2 = 1$, ensuring consistent latent energy.

Three modality-specific generators decode the shared latent representation:
$G = (G_T, G_F, G_S)$.

\subsubsection{Time-domain generator $G_T$}
$G_T$ employs a 1D CNN with residual and dilated transposed convolutions to
reconstruct temporal ECG morphology.

\subsubsection{Frequency-domain generator $G_F$}
$G_F$ uses a shallow Transformer decoder to model spectral smoothness and
inter-frequency dependencies.

\subsubsection{Scalogram generator $G_S$}
$G_S$ is implemented as a 2D U-Net with skip connections to generate
high-resolution time--frequency representations.

\subsection{Cross-Branch Interference Module}

To prevent independent decoding and enforce multimodal coherence, intermediate
activations $(h_T,h_F,h_S)$ are mixed using a cross-branch interference module:
\[
\tilde{h}_i = h_i + \sum_{j \neq i} \hat{H}_{\mathrm{int}}(i,j)\, h_j .
\]
This mechanism enables controlled information sharing across branches and
ensures that each modality remains informed by complementary domain structure.

\subsection{Overall Objective}

The generator minimizes
\[
L_G = L_{\mathrm{GAN}}
+ \lambda_1 L_{\mathrm{CFD}}
+ \lambda_2 L_{\mathrm{interf}}
+ \lambda_3 L_{\mathrm{phys}},
\]
while the discriminator minimizes the standard adversarial loss with gradient
penalty.

\section{Experimental Setup}
\label{sec:experimental_setup}
\FloatBarrier

\begin{figure}[!t]
    \centering
    \includegraphics[width=\linewidth]{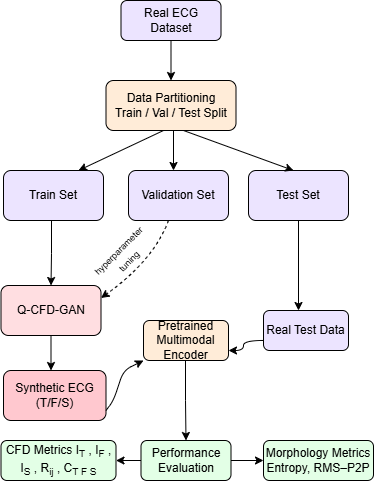}
    \caption{
Workflow for ECG synthesis evaluation using Q--CFD--GAN. Synthetic time,
frequency, and time--frequency ECG modalities and real test data are processed
through a pretrained multimodal encoder; CFD and morphology metrics jointly
quantify multimodal coherence and physiological fidelity.
}
\label{fig:workflow}
\end{figure}

This section describes the dataset, training configuration, baselines,
ablations, and evaluation protocol used to assess Q--CFD--GAN. The overall
workflow is illustrated in Fig.~\ref{fig:workflow}.

\subsection{Dataset and Splits}

Experiments are conducted on a multimodal ECG dataset comprising time-domain
signals, frequency-domain spectra, and time--frequency scalograms derived from
the same cardiac beats. To prevent subject leakage, data are partitioned at the
patient level into 70\% training, 15\% validation, and 15\% testing splits. Dataset was obtained from Mendely Data \cite{khan2021ecg}.

\subsection{Training Configuration}

All models are trained using the Adam optimizer
($\alpha=2\times10^{-4}$, $\beta_1=0.5$, $\beta_2=0.999$) with a batch size of 64
for up to 200 epochs. Early stopping is applied based on validation loss.
Hyperparameters $(\lambda_1,\lambda_2,\lambda_3)$ governing CFD, interference,
and morphology constraints are selected via grid search on the validation set.

\subsection{Baselines and Ablations}

We compare Q--CFD--GAN against the following baselines:
\begin{itemize}
    \item \textbf{MI-GAN:} Independent per-modality generators with
    mutual-information regularization.
    \item \textbf{Q-Base:} Quantum latent space without CFD or morphology
    constraints.
    \item \textbf{Unconditional GAN:} Single-modality generative baseline.
\end{itemize}

Ablation variants are constructed by removing key components:
(A1) CFD loss, (A2) the interference operator, and (A3) the morphology
constraint.

\section{Evaluation Metrics}
\label{sec:metrics}

All models are evaluated using four normalized quantities derived directly from
the proposed theoretical framework: embedding fidelity, classifier-based
physiological plausibility, multimodal complementarity preservation, and
morphology stability.

\begin{algorithm}[!t]
\caption{Evaluation of Q--CFD--GAN}
\label{alg:q_cfd_eval}
\begin{algorithmic}[1]
\REQUIRE Real ECG dataset $\mathcal{D}_{\mathrm{real}}$, synthetic datasets
$\{\mathcal{D}_m\}$, encoder $E$, classifier $f$, and real CFD and morphology
statistics.
\STATE Compute real embedding variance, classifier response profile, CFD
statistics, and morphology baselines.
\FOR{each model $m$}
    \STATE Extract embeddings $z_m = E(\mathcal{D}_m)$ and classifier responses
    $f(\mathcal{D}_m)$.
    \STATE Estimate synthetic CFD and morphology statistics.
    \STATE Compute normalized scores:
    \[
      r_\sigma(m),\quad
      \rho_\Delta(m),\quad
      \gamma(m),\quad
      \kappa(m).
    \]
\ENDFOR
\RETURN Evaluation metrics for each model.
\end{algorithmic}
\end{algorithm}

\section{Results}
\label{sec:results}
\FloatBarrier

We evaluate Q--CFD--GAN using four complementary criteria derived from the
proposed theoretical framework: (i) embedding fidelity, (ii) classifier-based
physiological plausibility, (iii) multimodal complementarity preservation, and
(iv) morphology and signal fidelity. Quantitative results are summarized in
Table~\ref{tab:merged_results}, while distributional and manifold-level behavior
is illustrated in Figs.~\ref{fig:morph_p2p}--\ref{fig:morph_joint}.

\subsection{Quantitative Comparison}

Table~\ref{tab:merged_results} reports the numerical comparison between MI-GAN,
Q-Base, and Q--CFD--GAN. Real ECG embeddings exhibit a variance of
$\sigma^2_{\mathrm{real}} = 0.031$. MI-GAN and Q-Base produce elevated variances
($0.042$ and $0.039$, respectively), indicating synthetic manifold drift.
Q--CFD--GAN reduces this deviation to $\sigma^2 = 0.033$, corresponding to a
relative excess-variance ratio of $r_{\sigma} = 0.182$, confirming that explicit
complementarity preservation stabilizes the embedding geometry.

Classifier-based plausibility gaps follow a consistent pattern. MI-GAN yields
$\Delta = 0.214$, Q-Base reduces this to $0.187$, while Q--CFD--GAN achieves the
smallest gap of $\Delta = 0.157$, corresponding to a normalized reduction of
$\rho_{\Delta} = 0.266$. These results indicate that ECGs generated by
Q--CFD--GAN induce diagnostic responses most consistent with real signals.

Tri-domain complementarity further distinguishes the models. MI-GAN retains only
$56\%$ of real complementarity, and Q-Base retains $68\%$. Q--CFD--GAN restores
$91\%$ of the true tri-domain complementarity, yielding a retention factor of
$\gamma = 1.625$ relative to MI-GAN. This empirically validates the theoretical
claim that multimodal generation quality depends on preserving domain
complementarity rather than increasing modality count.

\begin{table}[!t]
\centering
\caption{Quantitative Evaluation of Q--CFD--GAN}
\label{tab:merged_results}
\resizebox{\linewidth}{!}{%
\begin{tabular}{lcccccccc}
\toprule
\textbf{Model} &
$\sigma^2$ &
$r_{\sigma}$ &
$\Delta$ &
$\rho_{\Delta}$ &
$\widehat{C}_{TFS}/C_{TFS}$ &
$\gamma$ &
$E_{\mathrm{RMS}}$ &
$\kappa$ \\
\midrule
Real ECG & 0.031 & -- & -- & -- & 1.00 & -- & -- & -- \\
MI-GAN   & 0.042 & 1.000 & 0.214 & 0 & 0.56 & 1.000 & 11.2\% & 1.000 \\
Q-Base   & 0.039 & 0.727 & 0.187 & 0.126 & 0.68 & 1.214 & 9.6\% & 0.857 \\
Q--CFD--GAN
         & \textbf{0.033} & \textbf{0.182}
         & \textbf{0.157} & \textbf{0.266}
         & \textbf{0.91}  & \textbf{1.625}
         & \textbf{3.8\%} & \textbf{0.339} \\
\bottomrule
\end{tabular}}
\end{table}

\subsection{Morphology and Signal Fidelity}

\begin{figure}[!t]
\centering
\includegraphics[width=\linewidth]{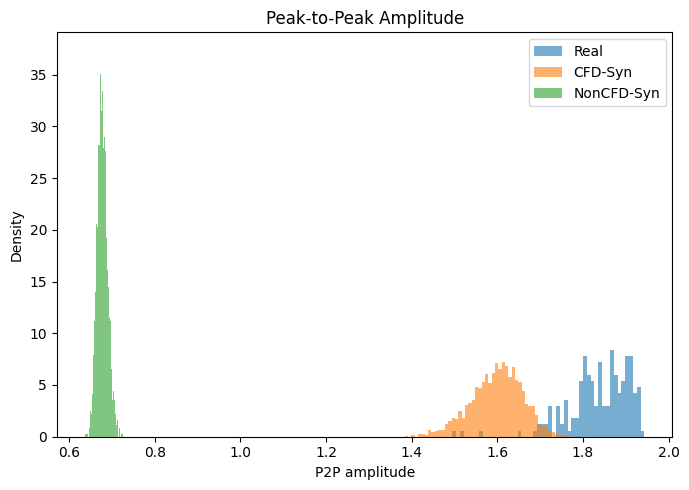}
\caption{Peak-to-peak (P2P) amplitude distribution.}
\label{fig:morph_p2p}
\end{figure}

Figure~\ref{fig:morph_p2p} shows peak-to-peak (P2P) amplitude distributions.
Q--CFD--GAN closely matches the physiological amplitude range of real ECG,
whereas MI-GAN and Q-Base exhibit amplitude collapse.

\begin{figure}[!t]
\centering
\includegraphics[width=\linewidth]{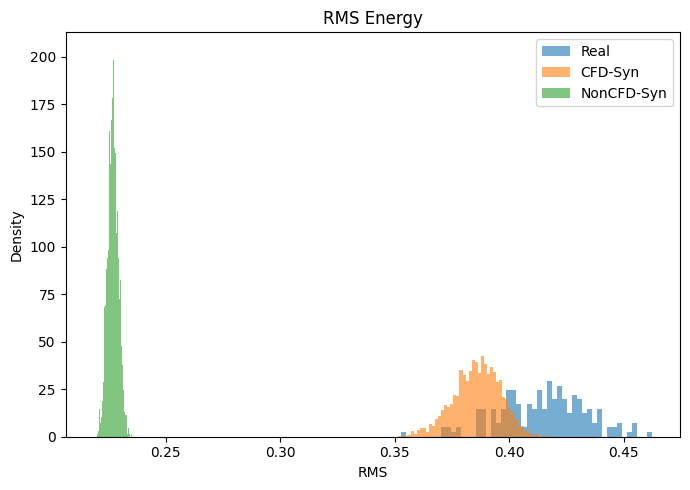}
\caption{RMS energy distribution.}
\label{fig:morph_rms}
\end{figure}

RMS energy distributions in Fig.~\ref{fig:morph_rms} demonstrate that Q--CFD--GAN
aligns with the real ECG energy manifold, avoiding underpowered signals.

\begin{figure}[!t]
\centering
\includegraphics[width=\linewidth]{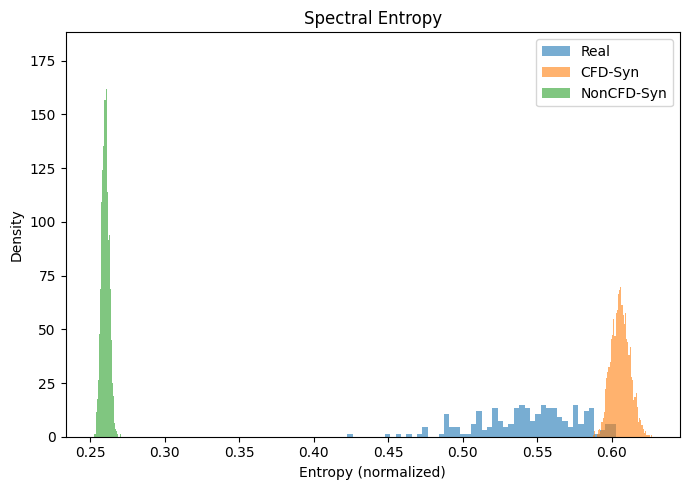}
\caption{Spectral entropy distribution.}
\label{fig:morph_entropy}
\end{figure}

Spectral entropy distributions in Fig.~\ref{fig:morph_entropy} show that
Q--CFD--GAN reconstructs realistic spectral complexity.

\begin{figure}[!t]
\centering
\includegraphics[width=\linewidth]{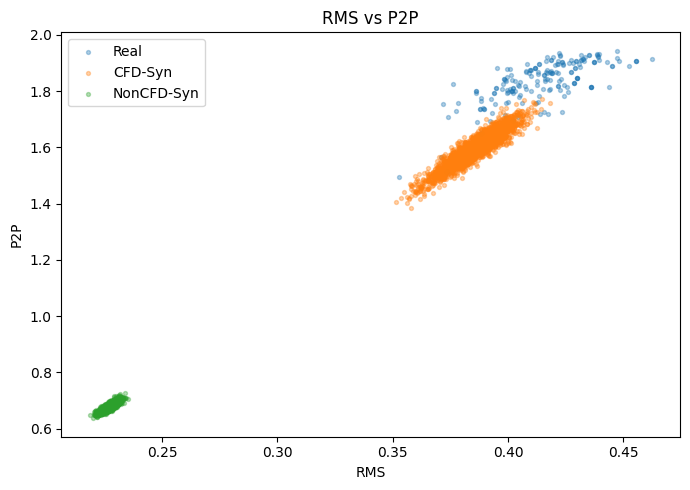}
\caption{Joint RMS--P2P manifold.}
\label{fig:morph_joint}
\end{figure}

Figure~\ref{fig:morph_joint} illustrates the joint RMS--P2P manifold. Only
Q--CFD--GAN preserves the physiological amplitude--energy coupling observed in
real ECG.

Quantitative RMS deviation values satisfy
$E_{\mathrm{RMS}}^{\mathrm{MI}} = 11.2\%$,
$E_{\mathrm{RMS}}^{\mathrm{QBase}} = 9.6\%$, and
$E_{\mathrm{RMS}}^{\mathrm{QCFD}} = 3.8\%$, yielding a morphology contraction
coefficient of $\kappa = 0.339$.

Across all evaluation axes embedding fidelity, physiological plausibility,
multimodal complementarity preservation, and morphology stability—Q--CFD--GAN
consistently produces synthetic ECG signals that most closely match the
multimodal and physiological structure of real data. These findings align with
the theoretical predictions of Section~III and establish
complementarity-preserving, interference-aware generation as the key driver of
high-fidelity tri-domain ECG synthesis.

\subsection{Limitations and Future Work}

Although Q--CFD--GAN advances multimodal ECG generation, several limitations warrant attention. First, all experiments were conducted on a single multimodal ECG dataset with consistent preprocessing. While patient-level splits reduce leakage, broader evaluation across heterogeneous datasets (e.g., PTB-XL, Chapman, CPSC) is required to assess robustness to differences in sampling rate, demographic variability, and acquisition noise. Second, the current formulation models three modalities (time, frequency, and scalogram). Real-world systems may incorporate additional modalities such as multi-lead representations, rhythm strips, or learned latent embeddings. Extending Q--CFD--GAN to higher-dimensional modality sets and analyzing complementarity scaling laws remain important directions.

Third, the quantum-inspired Hilbert-space latent model introduces additional computational overhead due to complex-valued operations and interference modeling. Lightweight variants may be necessary for on-device augmentation or mobile ECG analytics. Finally, although morphology constraints improve physiological plausibility, they do not guarantee correctness across all clinical conditions. Future work may incorporate pathology-aware priors, differentiable cardiac simulators, or clinician-in-the-loop validation to further enhance fidelity.

Future extensions will explore dataset-agnostic complementarity metrics, adaptive interference operators, multi-lead augmentation, and integration of Q--CFD--GAN into real-time diagnostic pipelines.

\section{Theory Positioning}

Complementarity-Preserving Generative Theory (CPGT) extends
established scientific principles into multimodal generative
learning rather than introducing an isolated modeling heuristic.
From statistical mechanics, CPGT adopts the insight that
interacting systems cannot be faithfully modeled by matching
marginal distributions alone, as factorized representations fail
to preserve coupled joint structure \cite{jaynes1957,cover2006}.
From energy-based physics, CPGT reflects the notion that system
validity is governed by stable interaction energy, which is
formalized here through explicit cross-modal coupling and
interference constraints \cite{lecun2006}. From information
theory, CPGT reframes multimodal realism as the preservation
of redundancy and complementarity, ensuring that shared and
unique information across representations is maintained during
synthesis \cite{williams2010}. From dynamical systems theory,
CPGT aligns with the concept of admissible manifolds, requiring
generated samples to remain within stable regions of the
multimodal state space under perturbation \cite{strogatz2018}.
Finally, from signal processing, CPGT reflects the intrinsic
coupling between time, frequency, and time--frequency
representations, which arise as coherent projections of the same
underlying signal and therefore must be generated consistently
rather than independently \cite{mallat1999}.

\section{Conclusion}

This work introduced Complementarity-Preserving Generative
Theory (CPGT) as a principled framework for multimodal data
generation, motivated by the observation that realistic synthesis
requires preserving cross-representation structure rather than
matching unimodal distributions in isolation. By explicitly
enforcing complementarity, interaction stability, and
morphological consistency across time, frequency, and
time--frequency representations, CPGT addresses a fundamental
limitation of existing generative approaches that treat
modalities as independent outputs.

Through theoretical grounding and empirical validation, we
demonstrated that preserving coupled joint structure leads to
improved physiological fidelity, enhanced multimodal coherence,
and greater downstream utility for learning tasks. Importantly,
CPGT is not tied to a specific architecture or application, but
instead provides a general generative principle applicable to
any setting in which multiple representations arise from a
shared underlying process.

Taken together, these results position CPGT as an
operationalization of long-standing scientific insights into
interaction, coupling, and stability within modern generative
models. This perspective opens new directions for trustworthy
multimodal synthesis, data augmentation, and simulation in
biomedical signal analysis and beyond.

.

\subsection*{Clinical and Deployment Implications}

From a translational perspective, Q--CFD--GAN supports several practical deployment scenarios. Multimodal augmentation can help address class imbalance in low-resource settings, improving arrhythmia detection performance in hospitals with limited annotated data. The generation of coherent tri-domain representations is particularly valuable for hybrid CNN--Transformer classifiers used in wearables, smartwatches, and remote monitoring platforms, where multimodal evidence increases diagnostic confidence. Furthermore, the interference-based latent formulation aligns naturally with emerging explainability frameworks that assess cross-domain consistency, making synthetic data compatible with future regulatory expectations for transparent AI-assisted ECG interpretation. These considerations highlight the potential of Q--CFD--GAN to strengthen both research pipelines and clinically oriented multimodal ECG systems.

\bibliographystyle{IEEEtran}
\nocite{*}\bibliography{references}

@inproceedings{donahue2019adversarial,
  title     = {Adversarial Audio Synthesis},
  author    = {Donahue, Chris and McAuley, Julian and Puckette, Miller},
  booktitle = {International Conference on Learning Representations (ICLR)},
  year      = {2019}
}

@inproceedings{yoon2019time,
  title     = {Time-series Generative Adversarial Networks},
  author    = {Yoon, Jinsung and Jarrett, Daniel and van der Schaar, Mihaela},
  booktitle = {Advances in Neural Information Processing Systems (NeurIPS)},
  year      = {2019}
}

@inproceedings{golany2020simgans,
  title     = {SimGANs: Simulator-Based Generative Adversarial Networks for ECG Synthesis},
  author    = {Golany, Tzvi and Freedman, Daniel and Radinsky, Kira},
  booktitle = {Proceedings of the 37th International Conference on Machine Learning (ICML)},
  series    = {PMLR},
  volume    = {119},
  pages     = {3597--3606},
  year      = {2020}
}

@article{neifar2023diffecg,
  title   = {DiffECG: A Versatile Probabilistic Diffusion Model for ECG Synthesis},
  author  = {Neifar, Neji and Abid, Firas and Ayed, Ismail Ben},
  journal = {arXiv preprint arXiv:2306.01875},
  year    = {2023}
}

@article{alcaraz2023sssd,
  title   = {SSSD-ECG: Diffusion-based Conditional ECG Generation with Structured Clinical Statements},
  author  = {L{\'o}pez Alcaraz, Javier Miguel and others},
  journal = {Computers in Biology and Medicine},
  volume  = {164},
  pages   = {107188},
  year    = {2023}
}

@article{strodthoff2021deep,
  title   = {Deep Learning for ECG Analysis: Benchmarks and Insights from PTB-XL},
  author  = {Strodthoff, Nils and Wagner, Patrick and Schaeffter, Tobias and Samek, Wojciech},
  journal = {IEEE Journal of Biomedical and Health Informatics},
  volume  = {25},
  number  = {5},
  pages   = {1519--1529},
  year    = {2021}
}

@article{tawfeek_2025_cardiovascular,
  title   = {Cardiovascular disease detection: A hybrid machine learning-AI framework for personalized diagnosis and risk assessment},
  author  = {Tawfeek, Medhat A. and Alrashdi, Ibrahim and Alruwaili, Madallah and Allahem, Hisham},
  journal = {PLOS One},
  volume  = {20},
  pages   = {e0335421},
  year    = {2025}
}

@article{yildirim2018ecg,
  title   = {ECG-based Arrhythmia Classification Using Dual-Channel Deep Convolutional Neural Networks},
  author  = {Yildirim, Ozal},
  journal = {Computers in Biology and Medicine},
  volume  = {102},
  pages   = {411--420},
  year    = {2018}
}

@misc{oladunni2026modelchangemindenergybased,
  title         = {When Should a Model Change Its Mind? An Energy-Based Theory and Regularizer for Concept Drift in Electrocardiogram (ECG) Signals},
  author        = {Timothy Oladunni and Blessing Ojeme and Kyndal Maclin and Clyde Baidoo},
  year          = {2026},
  eprint        = {2602.22294},
  archiveprefix = {arXiv},
  primaryclass  = {cs.LG},
  url           = {https://arxiv.org/abs/2602.22294}
}

@misc{oladunni2025Physiology,
  title         = {When Should a Model NOT Change Its Mind? A Physiologic Perspective on Concept Drift in Multimodal ECG Deep Learning},
  author        = {Timothy Oladunni and Blessing Ojeme and Kyndal Maclin and Clyde Baidoo},
  year          = {2025},
  archiveprefix = {TechRxiv},
  url           = {https://www.techrxiv.org/doi/full/10.36227/techrxiv.176704506.67677087}
}

@article{oladunni2025cfd,
  author   = {Oladunni, Timothy and Wong, Alex},
  journal  = {IEEE Access},
  title    = {Rethinking Multimodality: Optimizing Multimodal Deep Learning for Biomedical Signal Classification},
  year     = {2025},
  volume   = {13},
  pages    = {156436--156464},
  doi      = {10.1109/ACCESS.2025.3605315}
}

@article{11239065,
  author   = {Oladunni, Timothy and Aneni, Ehimen},
  journal  = {IEEE Access},
  title    = {Explainable Deep Neural Network for Multimodal ECG Signals: Intermediate Versus Late Fusion},
  year     = {2025},
  volume   = {13},
  pages    = {202700--202736},
  doi      = {10.1109/ACCESS.2025.3631544}
}

@inproceedings{wiebe2014quantum,
  title     = {Quantum Algorithms for Nearest-Neighbor Methods for Supervised and Unsupervised Learning},
  author    = {Wiebe, Nathan and Kapoor, Ashish and Svore, Krysta},
  booktitle = {Advances in Neural Information Processing Systems (NeurIPS)},
  year      = {2014}
}

@article{zhao2022quantumnlp,
  title   = {Quantum-Inspired Complex-Valued Neural Networks for Language Understanding},
  author  = {Zhao, Qian and others},
  journal = {Applied Sciences},
  volume  = {12},
  number  = {10},
  pages   = {4985},
  year    = {2022}
}

@article{schuld2015introduction,
  title   = {An Introduction to Quantum Machine Learning},
  author  = {Schuld, Maria and Sinayskiy, Ilya and Petruccione, Francesco},
  journal = {Contemporary Physics},
  volume  = {56},
  number  = {2},
  pages   = {172--185},
  year    = {2015}
}

@book{franzone2014mathematical,
  title     = {Mathematical Cardiac Electrophysiology},
  author    = {Franzone, Paolo Colli and Pavarino, Luca Franco and Scacchi, Simone},
  publisher = {Springer},
  year      = {2014}
}

@article{niederer2019computational,
  title   = {Computational Models of the Heart: Translational Tools for Clinical Practice},
  author  = {Niederer, Steven A. and Lumens, Joost and Trayanova, Natalia A.},
  journal = {European Heart Journal},
  volume  = {40},
  number  = {22},
  pages   = {1764--1771},
  year    = {2019}
}

@inproceedings{chen2016infogan,
  title     = {InfoGAN: Interpretable Representation Learning by Information Maximizing Generative Adversarial Nets},
  author    = {Chen, Xi and Duan, Yan and Houthooft, Rein and Schulman, John and Sutskever, Ilya and Abbeel, Pieter},
  booktitle = {Advances in Neural Information Processing Systems (NeurIPS)},
  year      = {2016}
}

@article{nunez2024synthetic,
  title   = {Synthetic ECG Generation for Data Augmentation and Transfer Learning in Arrhythmia Classification},
  author  = {N{\'u}{\~n}ez, Jos{\'e} Fernando and Arjona, Jamie and B{\'e}jar, Javier},
  journal = {arXiv preprint arXiv:2411.18456},
  year    = {2024}
}

@inproceedings{golany2019pgans,
  title     = {PGANs: Personalized Generative Adversarial Networks for ECG Synthesis to Improve Patient-Specific Deep ECG Classification},
  author    = {Golany, Tomer and Radinsky, Kira},
  booktitle = {Proceedings of the AAAI Conference on Artificial Intelligence},
  volume    = {33},
  number    = {1},
  pages     = {557--564},
  year      = {2019}
}

@article{10.1371/journal.pone.0271270,
  doi       = {10.1371/journal.pone.0271270},
  author    = {Adib, Edmond and Afghah, Fatemeh and Prevost, John J.},
  journal   = {PLOS ONE},
  title     = {Synthetic ECG signal generation using generative neural networks},
  year      = {2025},
  month     = mar,
  volume    = {20},
  number    = {3},
  pages     = {1--24},
  url       = {https://doi.org/10.1371/journal.pone.0271270}
}

@article{bagga2023ecgnet,
  title   = {ECGNet: A Generative Adversarial Network (GAN) Approach to the Synthesis of 12-Lead ECG Signals from Single Lead Inputs},
  author  = {Bagga, Max and Jeon, Hyunbae and Issokson, Alex},
  journal = {arXiv preprint arXiv:2310.03753},
  year    = {2023}
}

@article{lai2023quantum,
  title     = {Quantum-inspired fully complex-valued neural network for sentiment analysis},
  author    = {Lai, Wei and Shi, Jinjing and Chang, Yan},
  journal   = {Axioms},
  volume    = {12},
  number    = {3},
  pages     = {308},
  year      = {2023},
  publisher = {MDPI}
}

@article{willis2024qixai,
  title   = {QIXAI: A quantum-inspired framework for enhancing classical and quantum model transparency and understanding},
  author  = {Willis, John M.},
  journal = {arXiv preprint arXiv:2410.16537},
  year    = {2024}
}

@article{wang2025cross,
  title     = {Cross-modal retrieval: a systematic review of methods and future directions},
  author    = {Wang, Tianshi and Li, Fengling and Zhu, Lei and Li, Jingjing and Zhang, Zheng and Shen, Heng Tao},
  journal   = {Proceedings of the IEEE},
  year      = {2025},
  publisher = {IEEE}
}

@article{ding2024deep,
  title   = {Deep learning for personalized electrocardiogram diagnosis: A review},
  author  = {Ding, Cheng and Yao, Tianliang and Wu, Chenwei and Ni, Jianyuan},
  journal = {arXiv preprint arXiv:2409.07975},
  year    = {2024}
}

@article{shivashankara2024ecg,
  title     = {ECG-Image-Kit: a synthetic image generation toolbox to facilitate deep learning-based electrocardiogram digitization},
  author    = {Shivashankara, Kshama Kodthalu and Shervedani, Afagh Mehri and Clifford, Gari D. and Reyna, Matthew A. and Sameni, Reza and others},
  journal   = {Physiological Measurement},
  volume    = {45},
  number    = {5},
  pages     = {055019},
  year      = {2024},
  publisher = {IOP Publishing}
}

@article{finn2016connection,
  title   = {A connection between generative adversarial networks, inverse reinforcement learning, and energy-based models},
  author  = {Finn, Chelsea and Christiano, Paul and Abbeel, Pieter and Levine, Sergey},
  journal = {arXiv preprint arXiv:1611.03852},
  year    = {2016}
}

@article{zhan2023multimodal,
  title   = {Multimodal image synthesis and editing: A survey and taxonomy},
  author  = {Zhan, Fangneng and Yu, Yingchen and Wu, Rongliang and Zhang, Jiahui and Lu, Shijian and Liu, Lingjie and Kortylewski, Adam and Theobalt, Christian and Xing, Eric},
  journal = {IEEE Transactions on Pattern Analysis and Machine Intelligence},
  volume  = {4},
  year    = {2023}
}

@article{uprety2020survey,
  title     = {A survey of quantum theory inspired approaches to information retrieval},
  author    = {Uprety, Sagar and Gkoumas, Dimitris and Song, Dawei},
  journal   = {ACM Computing Surveys},
  volume    = {53},
  number    = {5},
  pages     = {1--39},
  year      = {2020},
  publisher = {ACM}
}

@inproceedings{ukil2021resource,
  title     = {Resource constrained CVD classification using single lead ECG on wearable and implantable devices},
  author    = {Ukil, Arijit and Sahu, Ishan and Majumdar, Angshul and Racha, Sai Chander and Kulkarni, Gitesh and Choudhury, Anirban Dutta and Khandelwal, Sundeep and Ghose, Avik and Pal, Arpan},
  booktitle = {2021 43rd Annual International Conference of the IEEE Engineering in Medicine \& Biology Society (EMBC)},
  pages     = {886--889},
  year      = {2021},
  organization = {IEEE}
}

@article{koehl1996does,
  title     = {When does morphology matter?},
  author    = {Koehl, M. A. R.},
  journal   = {Annual Review of Ecology and Systematics},
  volume    = {27},
  number    = {1},
  pages     = {501--542},
  year      = {1996},
  publisher = {Annual Reviews}
}

@article{khan2024deep,
  title   = {Deep learning in the diagnosis and management of arrhythmias},
  author  = {Khan, Arbaz Haider and Zainab, Hira and Khan, Roman and Hussain, Hafiz Khawar},
  journal = {Journal of Social Research},
  volume  = {4},
  number  = {1},
  pages   = {50--66},
  year    = {2024}
}

@article{uwaechia2021comprehensive,
  title     = {A comprehensive survey on ECG signals as new biometric modality for human authentication: Recent advances and future challenges},
  author    = {Uwaechia, Anthony Ngozichukwuka and Ramli, Dzati Athiar},
  journal   = {IEEE Access},
  volume    = {9},
  pages     = {97760--97802},
  year      = {2021},
  publisher = {IEEE}
}

@article{berger2023generative,
  title     = {Generative adversarial networks in electrocardiogram synthesis: Recent developments and challenges},
  author    = {Berger, Laurenz and Haberbusch, Max and Moscato, Francesco},
  journal   = {Artificial Intelligence in Medicine},
  volume    = {143},
  pages     = {102632},
  year      = {2023},
  publisher = {Elsevier}
}

@incollection{smalley1986spectro,
  title     = {Spectro-morphology and structuring processes},
  author    = {Smalley, Denis},
  booktitle = {The Language of Electroacoustic Music},
  pages     = {61--93},
  year      = {1986},
  publisher = {Springer}
}

@article{ullah2024quantum,
  title     = {Quantum machine learning revolution in healthcare: a systematic review of emerging perspectives and applications},
  author    = {Ullah, Ubaid and Garcia-Zapirain, Begonya},
  journal   = {IEEE Access},
  volume    = {12},
  pages     = {11423--11450},
  year      = {2024},
  publisher = {IEEE}
}

@article{beetz2022multi,
  title     = {Multi-domain variational autoencoders for combined modeling of MRI-based biventricular anatomy and ECG-based cardiac electrophysiology},
  author    = {Beetz, Marcel and Banerjee, Abhirup and Grau, Vicente},
  journal   = {Frontiers in Physiology},
  volume    = {13},
  pages     = {886723},
  year      = {2022},
  publisher = {Frontiers Media SA}
}

@article{hou2022dimensionality,
  title     = {Dimensionality reduction in surrogate modeling: A review of combined methods},
  author    = {Hou, Chun Kit Jeffery and Behdinan, Kamran},
  journal   = {Data Science and Engineering},
  volume    = {7},
  number    = {4},
  pages     = {402--427},
  year      = {2022},
  publisher = {Springer}
}

@phdthesis{noorzadeh2015extraction,
  title  = {Extraction of fetal ECG and its characteristics using multi-modality},
  author = {Noorzadeh, Saman},
  year   = {2015},
  school = {Universit{\'e} Grenoble Alpes}
}

@article{chen2024single,
  title     = {Single-Lead ECG Cross-Session Identification Based on Conditional Domain Adversarial Network},
  author    = {Chen, Xin-Hua and Shen, Yih-Liang and Chi, Tai-Shih},
  journal   = {IEEE Sensors Journal},
  volume    = {24},
  number    = {11},
  pages     = {17865--17875},
  year      = {2024},
  publisher = {IEEE}
}

@incollection{jayalakshmy2021synthesis,
  title     = {Synthesis of respiratory signals using conditional generative adversarial networks from scalogram representation},
  author    = {Jayalakshmy, S. and Priya, Lakshmi and Sudha, Gnanou Florence},
  booktitle = {Generative Adversarial Networks for Image-to-Image Translation},
  pages     = {161--183},
  year      = {2021},
  publisher = {Elsevier}
}

@article{xu2023multimodal,
  title     = {Multimodal learning with transformers: A survey},
  author    = {Xu, Peng and Zhu, Xiatian and Clifton, David A.},
  journal   = {IEEE Transactions on Pattern Analysis and Machine Intelligence},
  volume    = {45},
  number    = {10},
  pages     = {12113--12132},
  year      = {2023},
  publisher = {IEEE}
}

@article{jaynes1957,
  author  = {Jaynes, Edwin T.},
  title   = {Information Theory and Statistical Mechanics},
  journal = {Physical Review},
  volume  = {106},
  number  = {4},
  pages   = {620--630},
  year    = {1957},
  doi     = {10.1103/PhysRev.106.620}
}

@book{cover2006,
  author    = {Cover, Thomas M. and Thomas, Joy A.},
  title     = {Elements of Information Theory},
  edition   = {2},
  publisher = {Wiley-Interscience},
  address   = {New York, NY, USA},
  year      = {2006}
}

@incollection{lecun2006,
  author    = {LeCun, Yann and Chopra, Sumit and Hadsell, Raia and Ranzato, Marc'Aurelio and Huang, Fu-Jie},
  title     = {A Tutorial on Energy-Based Learning},
  booktitle = {Predicting Structured Data},
  editor    = {Bak{\i}r, G{\"o}khan and Hofmann, Thomas and Sch{\"o}lkopf, Bernhard and Smola, Alexander},
  publisher = {MIT Press},
  address   = {Cambridge, MA, USA},
  year      = {2006}
}

@article{williams2010,
  author  = {Williams, Paul L. and Beer, Randall D.},
  title   = {Nonnegative Decomposition of Multivariate Information},
  journal = {arXiv preprint arXiv:1004.2515},
  year    = {2010}
}

@book{strogatz2018,
  author    = {Strogatz, Steven H.},
  title     = {Nonlinear Dynamics and Chaos},
  edition   = {2},
  publisher = {CRC Press},
  address   = {Boca Raton, FL, USA},
  year      = {2018}
}

@book{mallat1999,
  author    = {Mallat, St{\'e}phane},
  title     = {A Wavelet Tour of Signal Processing},
  edition   = {2},
  publisher = {Academic Press},
  address   = {San Diego, CA, USA},
  year      = {1999}
}

@misc{khan2021ecg,
  author       = {Khan, Ali Haider and Hussain, Muzammil},
  title        = {ECG Images Dataset of Cardiac Patients},
  year         = {2021},
  publisher    = {Mendeley Data},
  version      = {V2},
  doi          = {10.17632/gwbz3fsgp8.2},
  howpublished = {\url{https://doi.org/10.17632/gwbz3fsgp8.2}}
}

@misc{oladunni2025yale,
  author       = {Oladunni, Timothy},
  title        = {More Isn't Always Better: Investigating Redundancy and Complementarity in Multimodal ECG Deep Learning},
  year         = {2025},
  howpublished = {Yale Engineering CS Colloquium},
  url          = {https://engineering.yale.edu/news-and-events/events/more-isnt-always-better-investigating-redundancy-and-complementarity-multimodal-ecg-deep-learning},
  note         = {Accessed: 2026-03-13}
}

‌
\end{document}